\documentclass[12pt,preprint]{aastex}

\def\halpha{H$\alpha$}

\def\sqig{$\sim$}

\def\cts{counts~s$^{-1}$}
\def\ergs{ergs~s$^{-1}$}
\def\degrees{$^{\circ}$}

\def\source{4U\,2206+54}
\def\src{4U\,2206+54}
\def\halpha{H$\alpha$}

\begin{document}

\title{Swift/BAT and RXTE Observations of the Peculiar X-ray
Binary 4U 2206+54 - Disappearance of the 9.6 Day Modulation}  

\author{R.H.D. Corbet\altaffilmark{1},
C.B. Markwardt\altaffilmark{2,3}, and J. Tueller\altaffilmark{3}}

\altaffiltext{1}{Universities Space Research Association;
X-ray Astrophysics Laboratory, Mail Code 662,
NASA Goddard Space Flight Center, Greenbelt, MD 20771; 
corbet@milkyway.gsfc.nasa.gov.}
\altaffiltext{2}{Department of Astronomy, University of Maryland,
College Park, MD 20742.}
\altaffiltext{3}{Astroparticle Physics Laboratory,
Mail Code 661, NASA/Goddard Space Flight Center, Greenbelt, MD 20771.}

\begin{abstract}
Observations of the high-mass X-ray binary \src\ with the Swift Burst
Alert Telescope (BAT) do not show modulation at the previously
reported period of 9.6 days found from observations made with the
Rossi X-ray Timing Explorer (RXTE) All-Sky Monitor (ASM).  Instead,
the strongest peak in the power spectrum of the BAT light curve occurs
at a period of 19.25 $\pm$ 0.08 days, twice the period found with the
RXTE ASM.  The maximum of the folded BAT light curve is also delayed
compared to the maximum of the folded ASM light curve.  The most
recent ASM data folded on twice the 9.6 day period show similar
morphology to the folded BAT light curve. This suggests that the
apparent period doubling is a recent secular change rather than an
energy-dependent effect.  The 9.6 day period is thus not a permanent
strong feature of the light curve. We suggest that the orbital period
of \src\ may be twice the previously proposed value.

\end{abstract}
\keywords{stars: individual (\source) --- stars: neutron ---
X-rays: stars}

\section{Introduction}

The Uhuru X-ray source \src\ (Giacconi et al. 1972)
was identified with the optical counterpart
BD +53\degrees2790 by Steiner et al. (1984).
Optical photometry and the presence of a double
peaked \halpha\ line in the spectrum led Steiner
et al. (1984) to conclude that
BD +53\degrees2790
is a Be star. However, it has been argued that the optical
properties, in particular the behavior of the H$\alpha$
emission line,  show that this star
is actually a very peculiar active O type star (Negueruela \& Reig
2001, Rib\'o et al. 2006, Blay et al. 2006).

Saraswat \& Apparao (1992) reported the detection of X-ray pulsations
from \src\ with a period of 392s
from EXOSAT observations. However, Corbet \& Peele (2001)
argued that the period search technique used by Saraswat \& Apparao (1992)
was inappropriate in searching for coherent variability. Corbet \&
Peele (2001)
reanalyzed the same EXOSAT data set 
and did not find any evidence for pulsations.
In addition, Corbet \& Peele (2001) and Negueruela \& Reig (2001)
analyzed Rossi X-ray Timing Explorer (RXTE) Proportional Counter
Array (PCA) observations of \src\ made in 1997 and again failed to find
any pulsations.
The X-ray spectrum can be well fitted with a ``standard'' high-mass
X-ray binary spectrum (White, Swank, \& Holt 1983) of
a power law modified by a high-energy cut off (e.g. Corbet \& Peele
2001; Masetti et al. 2004). The lack of an iron line in the spectrum
(Negueruela \& Reig 2001; Masetti et al. 2004)
makes it similar to high-mass X-ray binaries containing
Be stars rather than supergiant donors which often
have significant emission near 6.4 keV (e.g. Nagase 1989). 
Despite the lack of pulsations, evidence for the presence of a neutron
star in \src\ has been claimed to come from X-ray spectroscopy.
Masetti et al. (2004) analyzed a  BeppoSAX observation of \src\
made in 1998 and found tentative evidence for the presence
of a cyclotron resonance absorption feature.
Torrej\'on et al. (2004) analyzed the same data set
together with RXTE observations made in 1997 and 2001
and also found possible evidence for the presence
of a cyclotron absorption feature around 30 keV.
Blay et al. (2005) also report the presence of an absorption
feature near 32 keV in spectra obtained with
INTEGRAL.

From an analysis of approximately 5.5 years of
observations of \src\ with the  RXTE
All-Sky Monitor (ASM)
Corbet \& Peele (2001) found modulation at a period
of 9.568 $\pm$ 0.004 days. If this modulation is due
to orbital variability then this would be one of the shortest
orbital periods known for a Be type system (Raguzova \&
Popov 2005).
Corbet \& Peele (2001)
looked for differences between ``odd'' and ``even'' outbursts
that would indicate that the underlying period was actually 
twice this value but did not find any evidence for this. 
Rib\'o et al. (2006) analyzed additional RXTE ASM data and
derived a refined period of 9.5591 $\pm$ 0.0007 days.
Overall, the X-ray light curve of \src\ differs from the majority
of Be/neutron star systems in that the source does not
display either periodic Type I outbursts that occur near
periastron passage separated
by quiescent intervals or less frequent
but larger Type II events which are not modulated
on the orbital period (e.g. Charles \& Coe 2003). 
Instead, although \src\ is very variable,
particularly on shorter timescales, 
it is essentially persistent on long timescales.

In summary, although initial results suggested that
\src\ was a member of the common Be/neutron star binary
class of objects, further observations have called this into question.
Optical spectroscopy has cast doubt on the Be star classification,
and the lack of pulsations means that the nature of the presumed
accreting compact object in the system is not yet 
securely determined.
Because of the lack of pulsations from \src\
Corbet \& Peele (2001) considered alternative models for \src\
in which the accreting object might be a white dwarf, black hole,
or the magnetosphere of a neutron star rather than the surface
of a neutron star.

In this paper we present an analysis of observations
of \src\ made with the Burst Alert Telescope (BAT) on board the
Swift satellite. These data do not show modulation
on the 9.6 day period but, unexpectedly, show a maximum
in the power spectrum at twice this period.
We update the ASM results and compare the folded BAT
and ASM light curves and show that recent ASM data show
similar behavior to that found with the BAT.
We find that the phase of the BAT maxima
in the light curve does not exactly coincide with either the
``odd'' or ``even'' outbursts on the 9.6 day period. 
We also present light curves of \src\ obtained with the RXTE
Proportional Counter Array (PCA) and High Energy Timing
Experiment (HEXTE) in 2001 which
cover orbital phases around the maximum in the folded BAT
light curve.
We make a comparison of
\src\ with other systems with somewhat similar properties.
While there is some similarity with GRO J2058+42, the overall
properties of
\src\ so far appear to be unique.

\section{Observations}

\subsection{Swift BAT}
The BAT is described in detail by Barthelmy et al. (2005) and data
reduction is described by Markwardt et al. (2005). In brief,
the BAT is a very wide field of view (1.4 sr half-coded)
hard X-ray telescope that utilizes
a 2.7 m$^2$ coded-aperture mask and a 0.52 m$^2$ CdZnTe detector array
divided into 32,768 detectors each with an area of 0.16 cm$^2$.
The pointing direction of the BAT is controlled by observations
using the narrow-field
XRT and UVOT instruments also onboard Swift which are primarily used
to study gamma-ray bursts and their afterglows. BAT observations
of X-ray sources are thus generally obtained in a serendipitous
and unpredictable fashion. The BAT
sky survey is therefore non-uniform and the
signal to noise level of each observation of a source depends
on the location of the source within the BAT field of view.
Typically the BAT observes 50\% to 80\%
of the sky each day.
The data considered here consists of individual ``snapshots'' which
have exposure times ranging between 150 to 2660 s with a mean
of 1000 s.
From these snapshots light curves are constructed in
4 energy bands: 14 to 24 (``A''), 24 to 50 (``B''), 
50 to 100 (``C''), and 100 to 195 (``D'') keV.
For the entire energy range the Crab produces approximately
0.045 \cts\ per fully illuminated  detector for an equivalent
on-axis source (hereafter abbreviated to \cts).
In each energy band the Crab produces 0.019, 0.018, 0.0077,
and 0.0012 \cts\ respectively, equivalent to
approximately 311, 295, 126, and 19.7 \cts\ for the entire array
after accounting for the 50\% obscuration by the mask.
The light curve used in our analysis 
spans the interval from MJD 53,352 to 53,633
(2004-12-13 to 2005-09-20).
The fractional live time (uncorrected for partial coding)
on \src\ is 15\%, that is
the average exposure per day is approximately 3.6 hours.
The
mean efficiency due to partial coding during the observations
is 58\%.

\subsection{RXTE ASM}

The RXTE ASM (Levine et al. 1996) consists of three similar
Scanning Shadow Cameras
which perform sets of 90 second pointed
observations (``dwells'') so as to cover \sqig80\% of the sky every
\sqig90 minutes.  
Light curves are available in three energy bands: 1.5 to 3.0 keV
(``A''), 3.0
to 5 keV (``B''), and 5 to 12 keV (``C'').
The Crab produces approximately 75.5
\cts\ in the ASM over the entire energy range and 
26.8 (A), 23.3 (B), and 25.4 (C) \cts\ in each energy band. 
Observations
of blank field regions away from the Galactic center suggest that
background subtraction may produce a systematic uncertainty of about 0.1
\cts\ (Remillard \& Levine 1997).
The ASM observations presented here cover the period
from MJD 50,087 to 53,866 (1996-01-05 to 2006-05-11).

\subsection{RXTE PCA and HEXTE}

In addition to the ASM, RXTE carries two coaligned pointed
instruments - the PCA and HEXTE.
The PCA is described in detail by Jahoda et al. (1996).  This
instrument is sensitive to X-rays with energies between 2 - 60 keV and
has
a total effective area of \sqig6500 cm$^2$.  The Crab produces 13,000
\cts\ for the entire PCA across the complete energy band. 
Full specifications of HEXTE are given by Rothschild et al. (1998).
This instrument has a \sqig1600 cm$^2$ collecting area
and covers the energy range 15 - 250 keV.

Observations of \src\ were obtained with the PCA and HEXTE between
MJD 52193 to 52201
(2001-10-11 to 2001-10-19). These observations were designed
to sample four phases of the assumed 9.6 day period. Spectral
results from these observations
are reported by Torrej\'on et al. (2004).
For our analysis we use the ``standard products'' light curves provided 
by the RXTE Guest Observer Facility 
\footnote{http://heasarc.gsfc.nasa.gov/docs/xte/recipes/stdprod\_guide.html}
(Boyd et al. 2001).
Although not intended for detailed analysis, these RXTE standard
products are sufficient to provide the general light curve
structure and are only used for that
purpose in this paper. PCA light curves from the ``Standard 2'' mode are
background subtracted using the appropriate model
for each gain epoch and have a time resolution of 16s.
In this paper we employ the 2 - 9 keV light curve.
Background subtracted HEXTE light curves have a time resolution
of 16s and are available for the energy ranges 15 - 30 keV,
30 - 60 keV, and 60 - 250 keV.

\section{Analysis and Results}

\subsection{Weighting Data Points in Power Spectra}
Due to various factors such as source location in
the detector field of view, exposure duration,
and non-uniform sky coverage, 
the light curves from both the BAT and ASM show considerable
variability in the statistical quality of individual
data points. When searching for periodic modulation
it is common to calculate power spectra. Two common ways
to calculate the power spectrum of a data set with non-uniform
error bars are to either weight the contribution of a data point
to the power spectrum (e.g. Scargle 1989) 
or to use uniform weighting.
The use of weighting is briefly discussed for RXTE ASM data in Corbet (2003)
where a weighting factor of $y_i/\sigma_i^2$ (``simple weighting'')
was used where $y_i$ is the value of a data point and $\sigma_i$
is its associated error.
Whether weighted or unweighted power spectra was preferred
was argued by Corbet (2003) to depend on the relative size of
the error bars compared to overall variability. 
If the scatter in data values is large compared to error
bar size then weighting by error bars can be inappropriate.
In addition,
it was noted by Corbet (2003) that
periodic or quasi-periodic changes in data quality, such
as caused by the precession period of RXTE's orbit,
may introduce
spurious signals in the power spectrum if weighting is used.
The effects of periodicities in data quality in ASM
light curves are also noted by Wen et al. (2006). 

For the Swift BAT we investigated the difference between
``weighted'' and ``unweighted'' power spectra for a number of
high-mass X-ray binaries with already well determined orbital 
and superorbital periods
that covered a range of count rates. As a quality parameter we
used the ratio of the height of the peak in the power spectrum
compared to the mean noise level. We then calculated the relative
peak heights measured in this way for weighted and unweighted
power spectra and this ratio is plotted against count rate in
Fig. 1.
We find that, as predicted, the benefits of weighting are strongly dependent
on source brightness, with weighting very beneficial
for faint sources, but detrimental for bright sources.
The count rate of \src\ lies in the transition region between weighting
being detrimental to beneficial. 
We next investigated an alternative weighting
scheme (``modified weighting'')
which incorporates the variance due to source variability.
The contribution of each data point to the power spectrum is
calculated as:
\begin{displaymath}
\frac{y_i}{((f \sigma_i)^2 + V_S)}
\end{displaymath}
where $V_S$, the estimated variance due to source variability,
is estimated by calculating the difference between the observed
variance and the mean square of the error bar sizes. i.e.:
\begin{displaymath}
V_S = \frac{\sum_{i=1}^N(\bar{y} - y_i)^2}{(N - 1)} 
 - \frac{\sum_{i=1}^N (f \sigma_i)^2}{N}
\end{displaymath}
The factor $f$ is a correction to nominal error
on each point, it is derived from
observations of sources expected to be constant such
as supernova remnants and galaxy clusters and is
set so that the $\chi^2$ value for constant count rate becomes
1. For the BAT the correction factor adopted was 1.2.
For cases where the calculated value of $V_S$ was found
to be less than
zero it was set equal to zero. 
A comparison of peak detection significance 
using this modified weighted scheme 
is also shown in Fig. 1. We note that
the modified weighting is comparable to simple weighting
for the weakest sources and is essentially equivalent to
unweighted power spectra for the brightest sources.
For sources with count rates comparable to
\src\ (IGR J19140+0951 and IGR J17252-3616) we find that modified weighting
results in period detections which are more significant than
if either simple weighting or no weighting is used. 
The modified weighting scheme was found to be at least as good
as, and often superior to, both unweighted and simply weighted
power spectra for all except the very faintest sources.
We therefore
employ this modified weighting for all power spectra in
this paper.

\subsection{BAT and ASM}

The light curves of \src\ obtained with the BAT and
ASM are shown in Figs. 2 and 3. \src\ is seen
to be very variable as reported in Corbet \& Peele (2001).
However, there is no long term trend in the source brightness,
and there are no obvious exceptionally bright or faint states
that the source remains in for any extended period of time.

The weighted mean count rate from \src\ for the entire BAT energy range is
3.3 $\times$ 10$^{-4}$ $\pm$ 1 $\times$10$^{-6}$ (statistical) \cts\
which corresponds to
7.3 mCrab. In the four energy bands the weighted mean count rates 
and their statistical errors are
A: 1.61 $\times$ 10$^{-4}$ $\pm$ 5 $\times$ 10$^{-6}$ \cts\ (8.5 mCrab), 
B: 1.22 $\times$ 10$^{-4}$ $\pm$ 5 $\times$ 10$^{-6}$ \cts\ (6.7 mCrab), 
C: 4.3 $\times$ 10$^{-5}$ $\pm$ 3 $\times$ 10$^{-6}$ \cts\ (5 mCrab), and
D: 7 $\times$ 10$^{-7}$  $\pm$ 2 $\times$ 10$^{-6}$ \cts\ ($<$ 2 mCrab).

The weighted mean
ASM count rate from \src\ is 0.365 $\pm$ 0.004 (statistical) \cts,
equivalent
to 4.9 mCrab. In each ASM energy band the weighted mean
count rates and statistical errors are
0.050 $\pm$ 0.002 (A), 0.079 $\pm$ 0.002 (B) and 0.178 $\pm$ 0.003 (C) \cts,
corresponding to
1.9, 3.4, and 7.0 mCrab respectively.  

In a light curve averaged over intervals of one day the
mean significance with which \src\ was detected in the BAT is
2.5$\sigma$. For the ASM the
mean significance with which \src\ was detected is
1.4$\sigma$ for the total light curve, and 1.14$\sigma$
for just the interval coinciding with the BAT observations.
The BAT thus detects \src\ at a higher average significance 
than does the ASM. However, due to the differences in the way
ASM and BAT observations are obtained the ASM provides more
uniform coverage than does the BAT.

We searched for periodic modulation from \src\ by first 
calculating the
power spectrum of the BAT light curve for the
entire 14 - 195 keV energy range and this is shown in the lower panel of
Fig. 4. No signal is present at the previously reported
9.6 day period, instead the strongest peak is at a period
of 19.28 days at a level of 19 times the
mean value of the power spectrum and \sqig6 times the local power
level. 
This period coincides
to better than 1\% with twice the 9.6 day period. We will use
the term ``2P'' hereafter to refer to the period of twice the originally
proposed orbital period near 9.6 days, i.e. 19.118 days.

We next calculated the power spectrum of each BAT
energy band
separately and the results for bands A and B are also shown
in Fig. 4. Band A shows multiple peaks
in the power spectrum including one near 19.28 days. 
and the power spectrum of
Band B has its strongest peak at 19.28 days
at a level of 19 times the mean power, and \sqig7 times
the local power. Apart from the peak at 19.28 days, 
there are less strong peaks in the B band power spectrum than
in the lower energy A band power spectrum. 
For neither spectrum do any of the peaks appear to be
related to each other as harmonics or aliases.
The BAT power spectra thus demonstrate the considerable
variability on a variety of timescales commented on
in previous work (Section 1). However, the clear
discrepancy from previous work is the lack of significant
modulation at 9.6 days.

We divided the B band light curve into two halves and
examined the power spectra of these separately. In both
cases we see peaks near 19 days and no significant peak
near 9.6 days.
Results are not
shown in Figure 4 for bands C and D due to the low count rates
in these bands.

We next calculated the power spectrum of the RXTE ASM to search
for modulation at a period of twice the previously reported value.
Although
the modified weighting described in Section 1 was employed with
an error bar modification factor, $f$, of 1.27
this procedure gave an estimated source
variance, $V_S$, of zero and hence was equivalent to
simple weighting. 
The resulting power spectrum (Fig. 5) shows a low amplitude
structured peak containing 2P with strongest
power at 19.27 days.

We next folded both the ASM and BAT data on 2P using
simple weighting and
the results are shown in Fig. 6. The ASM and BAT A and B band
data all show clear modulations. However, as expected from
the power spectra, the ASM light curve shows two nearly
equal peaks, whereas the BAT A and B bands shows only a single
broad peak.
We note that the peaks in the folded BAT light curves do
not appear to coincide in phase with either peak in the ASM
folded light curve.
In these folded light curves we also note the possible presence
of a brief flux increase in the folded BAT 14 - 24 keV light curve
near a phase of 0.75 (for a period of 2P).
Although possible corresponding features 
may also exist in the folded ASM light
curves and the 24 - 50 keV BAT light curve,
when the BAT light curve is divided into two halves we find that
this is not a persistent feature of the light curve.
The folded ASM overlap light curve also does not show any
significant feature near a phase of 0.75.
 
To further quantify periodic modulation we next
fitted sine waves
to the non-folded ASM (1.5 - 12 keV) and BAT (24 - 50 keV) light curves.
In these fits data points were weighted by their individual
errors. From the ASM dwell data we obtain a period of
9.5592 $\pm$ 0.0010 d. This is consistent with the value
of 9.5591 $\pm$ 0.0007 d obtained by Rib\'o et al. (2006).
The epoch of maximum flux was found to be MJD 53567.74 $\pm$ 0.23.
From the BAT data we obtained a period of 2P = 19.25 $\pm$ 0.08 d,
equivalent to P = 9.625 $\pm$ 0.04 d. The periods derived
from the ASM and BAT are thus consistent (when the
factor of 2 period difference is allowed for) to within 0.6\%, 
equivalent to a 1.6$\sigma$ difference.
To quantify any phase difference we next fixed the
BAT period
at twice the ASM derived value and allowed just the phase and
amplitude of the sine wave to float. 
We find that the zero crossing phases of the sine
fits to the ASM and BAT light curves are consistent within
the errors. Since the period found with the BAT is twice the 
length of that derived from the ASM,
the maximum of the BAT curve thus occurs later than
the corresponding maximum in the ASM light curve by
2.39 $\pm$ 0.5 days, equivalent to a phase
difference of 0.125 $\pm$ 0.026 for a period of 2P.
The results of the fits
to the non-folded light curves are also plotted over
the folded light curves in Fig. 6.

To investigate any changes in periodic modulation
in the ASM light curve we calculated power spectra
for subsets of the data using a sliding box. Power spectra
were calculated for two year long subsets of ASM data
starting from the first observations and incrementing the
start of the data subset by one year, resulting in a total
of 10 power spectra. The results of this procedure in a region
covering both 9.6 days  and 2P are
shown in Fig. 7. We note that the modulation at 
9.6 days is, in general, stronger in the first five years of ASM
observations. In the power spectra from the most recent subset
the strongest power in this frequency range is found at 2P,
consistent with the BAT results.
We also folded the data in the same subsets on a period of 2P
and these are shown in Fig. 8.
We note that the earlier subsets show the expected two peaks
while the last subset shows a profile comparable to
that seen with the BAT with a single maximum at similar phase.
In order to evaluate changes in the profile we first constructed
a template using the mean of the first two statistically independent
profiles (those shown in Fig. (a) and (c)). This template was
then subtracted from all profiles and these new profiles are shown
in Fig. 9. None of the profiles shows large systematic deviations
from the template profile with the exception of the
profile from the most recent observations (Fig. 9 (j))
where a pronounced feature near phase 0.3 is present.

\subsection{PCA and HEXTE}
The light curves obtained with the PCA and HEXTE
are shown in Fig. 10.
The modulation of the PCA light curve is consistent with the
folded ASM light curve. However, the HEXTE light curves,
which cover similar energy ranges to the BAT, do not show
similar behavior to the folded BAT light curve and appear
to follow the PCA light curve. In particular the HEXTE
light curves do not show a minimum near phase 0.5 which would be
expected if they followed the behavior seen with the BAT.
However, some caution should be taken in interpreting these
observations as \src\ shows significant apparently 
non-periodic behavior which can affect a short data set such as this.

\section{Discussion}

\subsection{Periodic Modulation and the Orbital Period}

The BAT light curve of \src\ shows modulation at a period
twice that previously derived from RXTE ASM data.
If this modulation is orbital in origin then it
suggests that the underlying orbital period may actually
be double the previously proposed value. The BAT light curve covers
282 days, equivalent to more than 14 ``2P'' cycles and thus
the apparent period doubling is unlikely to be a purely
stochastic effect caused by a small number of unusually bright maxima.
\src\ thus shows either an energy dependent number of X-ray maxima
per orbit, or else it underwent a transition between exhibiting two
maxima per orbit to a state of a single maximum per orbit.
The energy-dependent modulation hypothesis may be supported
by the shape of the folded BAT ``A'' band light curve which
has an appearance suggestive of a mix of the two types
of modulation. Thus a change in the nature of the periodic modulation may
occur in the range of 14 - 24 keV.
However, evidence that
argues against this hypothesis is: (i) The shape of the
HEXTE light curve (Fig. 7) does not follow the folded BAT
light curve (although non-periodic modulation presumably also
contributes to the observed light curve). 
(ii) The most recent ASM data folded on 2P
show the same morphology as the BAT light curve (Fig. 6). 

In principle, optical radial velocity measurements could
be used to determine the orbital period of \src\ unambiguously.
Blay (2006) presents such measurements obtained in 2002,
2003, and 2004. However, none of the individual campaigns
cover substantially more than 9.6 days. Blay (2006) reports that the power
spectrum of the complete set of radial velocity observations
does not show a clear peak at 9.6 days, but instead shows
a maximum at 14.89 $\pm$ 0.07 days which he interprets as
an artifact, perhaps related to noisy data. The currently
published radial velocity measurements are thus not yet
capable of independently determining the orbital period
of \src.

It could be possible that the periodic changes in
the high-energy emission from \src\ are not due to
orbital effects but are caused by super-orbital
modulation. For example,  a 30.7 day super-orbital period 
is seen the peculiar high-mass X-ray binary
2S 0114+650 (Farrell, Sood, \& O'Neill 2006)
in which accretion occurs from the wind of a supergiant B1 star
(Reig et al. 1996). 
However, it would still be difficult to
understand how a period doubling would occur.
If the periodic modulation in \src\ is indeed
a super-orbital effect then modulation at the orbital
period is apparently lacking.

Modulation of X-ray emission on the orbital period in
a high-mass X-ray binary may occur in several different ways
(e.g. White, Nagase, \& Parmar 1995 and references therein).
If the orbit is eccentric then enhanced emission can
occur at periastron passage.
For \src\ Rib\'o et al. (2006) attributed the periodicity
in the X-ray flux seen with the ASM to modulation caused  
by an eccentric (e = 0.15) orbit but the difference between
the BAT and ASM periods cannot be explained by this simple model alone.
If the primary star is large
compared to the orbital dimensions
(e.g. it is a supergiant) then eclipses are often observed.
Variable absorption may also be seen as the neutron star
moves in its orbit. In addition, if the orbital plane
is inclined to the plane of an equatorial circumstellar disk, 
such as
can be present in Be stars, then two outbursts per orbital period
may be possible when the neutron star passes through this disk.
If the disk shrinks or expands this could potentially change the
number of outbursts observed per orbit. Similar effects could
arise if there is a change in the angle between the orbit
of the neutron star and the disk.

We next summarize results on some X-ray binaries
which have the characteristics of either showing two maxima
per orbital cycle, energy dependent modulation, or phase
change in the epoch of flux maximum, and so might provide
an insight into the variability of \src.

The Be/neutron star binary GRO J2058+42 was initially
reported to have a 110 day period seen in CGRO BATSE (20 - 50 keV)
data. Strong outbursts every 110 days were accompanied by
weaker outbursts halfway between these. However,
RXTE ASM observations showed an underlying 55 day period
(Corbet, Peele, \& Remillard 1997). Wilson et al. (1998)
find that there is a clear energy difference in
the size of the outbursts detected and these authors
proposed that the orbital period of the system is 110 days
with outbursts occurring at both periastron and apastron passage.
The energy dependence of the underlying period in GRO J2058+42
has some apparent similarities to what is seen in \src.
However, for GRO J2058+42
no phase shift is observed. Wilson et al. (2005) also
report that pulse frequency changes seen during the outbursts
are consistent with a 55 day period. 

The supergiant system 4U 1907+09 was found by
Marshall \& Ricketts (1980) to exhibit two outbursts
per orbital cycle, separated by approximately 0.45
in phase. Marshall \& Ricketts noted that:
``the amplitudes of the secondary maxima are more variable
than those of the primary maxima ranging from being approximately
equal in flux down to insignificance.''
in 't Zand, Baykal, \& Strohmayer (1998) speculated
that the two flares per orbital
cycle might indicate that the wind from the supergiant
primary is not isotropic and the flares may occur
when the neutron star crosses the equatorial plane of
the primary where there is an enhanced wind. 
In the supergiant high-mass X-ray binary 4U 1223-62 (GX 301-2)
evidence has also been reported for the presence of
flares at apastron in its eccentric
41.5 day orbit in addition to the large flares that
occur shortly before periastron (Pravdo et al. 1995, Pravdo \& Ghosh 2001).
For 4U 1223-62 the size of the apastron flare has always been observed to
be much smaller than the periastron flare.

The phase difference between the ASM and BAT light curves
could also potentially be a secular or an energy dependent effect.
For the Be/neutron star binary EXO 2030+375, which has an
orbital period of 46 days, a temporary
change in the phase of maximum flux was seen to
occur in both RXTE ASM and BATSE data (Wilson et al. 2002)
with a difference of about 10 days. No period change accompanied
this temporary phase shift.
Corbet (2003) reported evidence for a 24 day period in the
{\em low-mass} X-ray binary GX 13+1 from RXTE ASM observations
and that the higher energy (5 - 12 keV) emission trailed
the lower energy (1.5 - 5 keV) by a phase difference
of 0.2 $\pm$ 0.03. 

Of the sources considered above, while none exhibits exactly the same
behavior shown by \src,
a possible similarity with GRO J2058+42 may exist in the apparent factor
of two difference in orbital periods found by two different
instruments at different times. 
However \src\ has several differences including the presence of
pulsations in GRO J2058+52 and the possible difference in the
nature of the mass donor. The variable secondary maxima
exhibited by 4U 1907+09 may also be similar to the behavior shown
by \src\ although there is again a difference in the
nature of the mass donor.

While the systems discussed here provide hints of how an
underlying orbital period at 2P might show itself as modulation
at half that period, it appears even more difficult to understand
how a \sqig9.6 day long orbital period could produce
modulation at twice that period.
We therefore suggest that it is more likely that the orbital
period of \src\ is 2P, that is, it is \sqig19.2 days.

If \src\ is at least similar to Be star systems, then
the correlation between orbital period and pulse period
for these objects (Corbet 1986) 
predicts a pulse period of a few seconds
for an orbital period of 19.2 days. However, a 19.2 day period
is still relatively short for a Be star system. The only
two objects currently known with shorter orbital periods are A0538-66
with an orbital period of 16.7 days
which has an exceptionally short pulse period of 69 ms (Skinner
et al. 1982)
and SAX J2103.5+4545 with an orbital period
of 12.7 days and a pulse period much longer than
expected from the pulse/orbital period
correlation of 358.6s (Baykal, Stark \& Swank 2000). These two
unusual sources may provide a hint that the
orbital period correlation could break down
at short orbital periods,

\subsection{Luminosity}

Corbet \& Peele (2001) considered the 2 - 10 keV luminosity of
\src\ at \sqig10$^{35}$ ergs s$^{-1}$
to be surprisingly low for the assumption of
a 9.6 day orbital period and a Be star mass donor. In contrast,
Rib\'o et al. (2006) determine the mass donor
not to be a Be star and thus they consider
the ``high'' luminosity in need of explanation due
to the relatively weak wind they predict compared to
the extensive circumstellar envelope that would
be present around a Be star.
With a 19.2 day orbital period the
luminosity would be expected to be lower than for a \sqig9.6
day period, due to the expected lower mass transfer rate,
and therefore not atypical for a Be star system.
The modeling of Rib\'o et al. (2006) would need to be
reevaluated for the new system parameters to determine whether
their model is still tenable.
For comparison, two Be star systems with orbital periods
close to 19.2 days are 4U 1901+03 (22.6 days; Galloway, Wang, \& Morgan 
2005) and
4U 0115+63 (24.3 days; Rappaport et al.
1978,  Bildsten et al. 1997). The luminosity ranges exhibited
by both these sources are very large and thus very different
from the variability seen so far from \src.
4U 0115+63 has been detected at luminosities
between  \sqig10$^{38}$ and \sqig10$^{38}$ \ergs\ (Campana et al.
2001 and references therein). Although the distance to 4U 1901+03 
is unknown, this source also exhibits
considerable variability. Galloway et al.  (2005) report 
RXTE PCA and ASM observations where  4U 1901+03 reached a peak
luminosity of  \sqig10$^{38}$ (d/10 kpc)$^2$ \ergs\ before
fading to a level $>\sim$ 100 times less.

\section{Conclusion}

The ASM and BAT light curves show that strong modulation
at a period of 9.6 days is not a persistent feature
of the light curve of \src.
The BAT light curve of \src\ shows a modulation at twice
the period previously found from RXTE ASM observations.
A detailed examination of the ASM light curve also demonstrates
a change in the nature of the modulation.
The orbital period of
\src\ might thus be twice the originally proposed value of
9.6 days. 
Although the mechanism causing the variable modulation in \src\
is not yet clear, a model where there is
an equatorial concentration of the wind
from the primary star and the equatorial and orbital
planes are offset may be able to explain
at least some of the observed properties.
It would be very valuable if a radial velocity orbit could
be obtained from optical observations as this should conclusively give
the value of the orbital period, show the relative orientation
of the binary components at the times of X-ray maxima, and give
the orbital eccentricity. 
If the period doubling is caused by a change in the circumstellar
environment in the system, then this may also show effects
in the optical spectrum such as the
strength or shape of the H$\alpha$ emission line. Continued
high-energy monitoring of \src\ with the BAT and/or ASM
will show how long the change to the period doubled state
continues which may assist in understanding its cause.

The nature of \src\ continues to be puzzling and the 
determination that the
9.6 day modulation is not a persistent feature of the light
curve adds to the peculiarities of the source. The
exceptional properties of \src\ may be related to
the unusual nature of the mass donor in the system.
\src\ may thus provide a new parameter regime where
models of wind accretion may be tested.

\acknowledgements
RHDC thanks Pere Blay for providing information
on his optical observations and Alan M. Levine for useful discussions on
the weighting of data points in power spectra and suggesting
the modified weighting technique.

\clearpage
\noindent
{\large\bf Figure Captions}

\figcaption[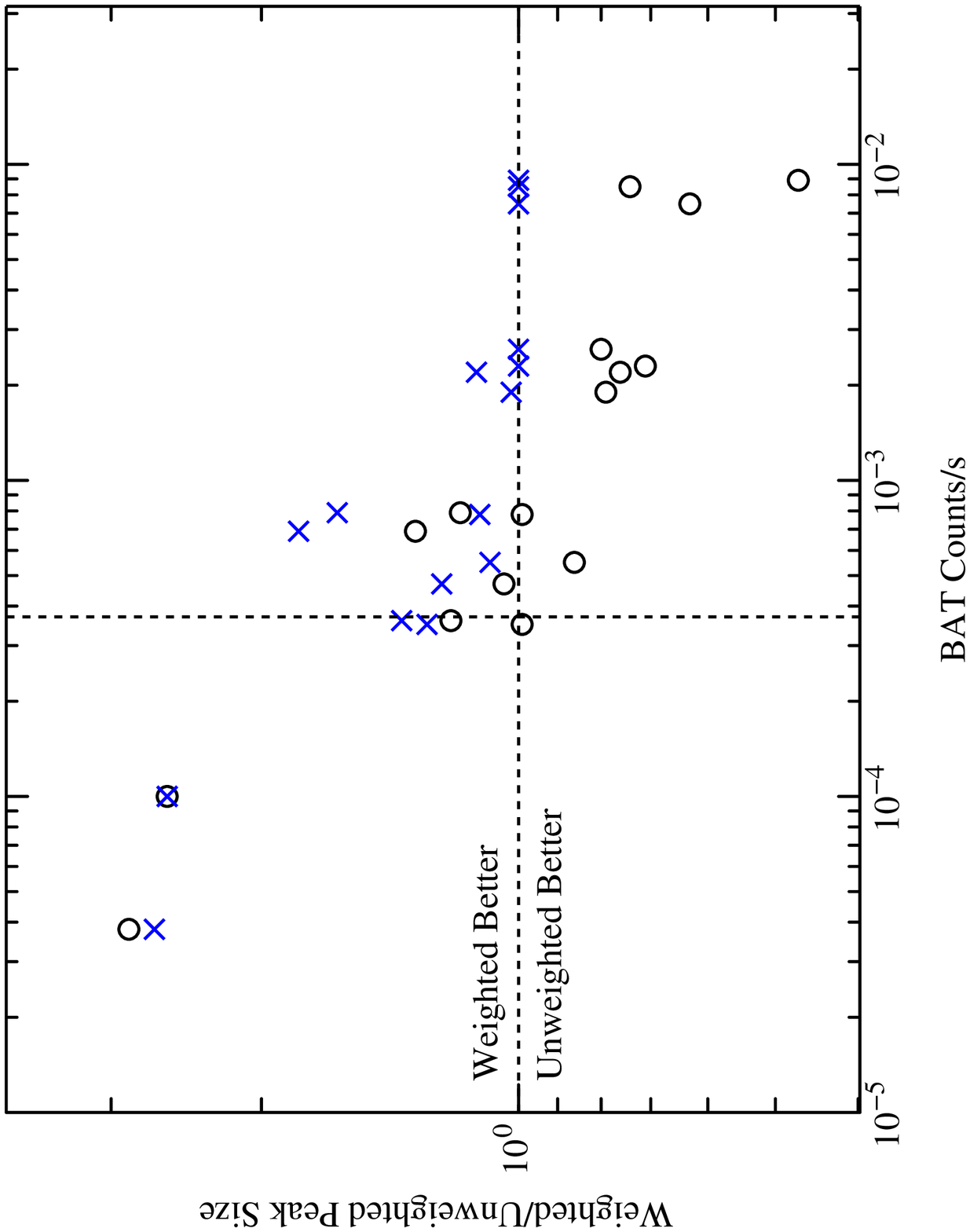]{Comparison of power spectrum weighting techniques
in detecting periodic modulation in BAT light curves.
The black circles show the ratio of signal peak sizes between
power spectra
weighted with simple weighting and no weighting,
and the blue crosses show
the ratio between ``modified weighting'' and no weighting.
Details are given in Section 3.1.
The horizontal dashed line shows the division between weighting
being better or worse for peak detection, i.e. whether the peak
in the weighted or unweighted power spectrum is larger.
The vertical dashed line shows the count rate of \src.
Systems on this plot and their mean count rates are:
X 0726-260 (3.8$\times$10$^{-5}$), IGR J11435-6109
(0.0001), IGR J19140+0951 (0.00035), EXO 1722-363 (0.00036),
XTE J1855-026 (0.00047), IGR J16320-4751 (0.00055),
X 1907+097 (0.00069), LMC X-4 (0.00078), 
X 1538-522 (0.00079), EXO 2030+375 (0.0019),
SMC X-1 (0.0022), Cen X-3 (0.0023), OAO 1657-415 (0.0026),
GX 301-2 (0.0075), X 1700-377 (0.0085), and Vela X-1 (0.0089).
}

\figcaption[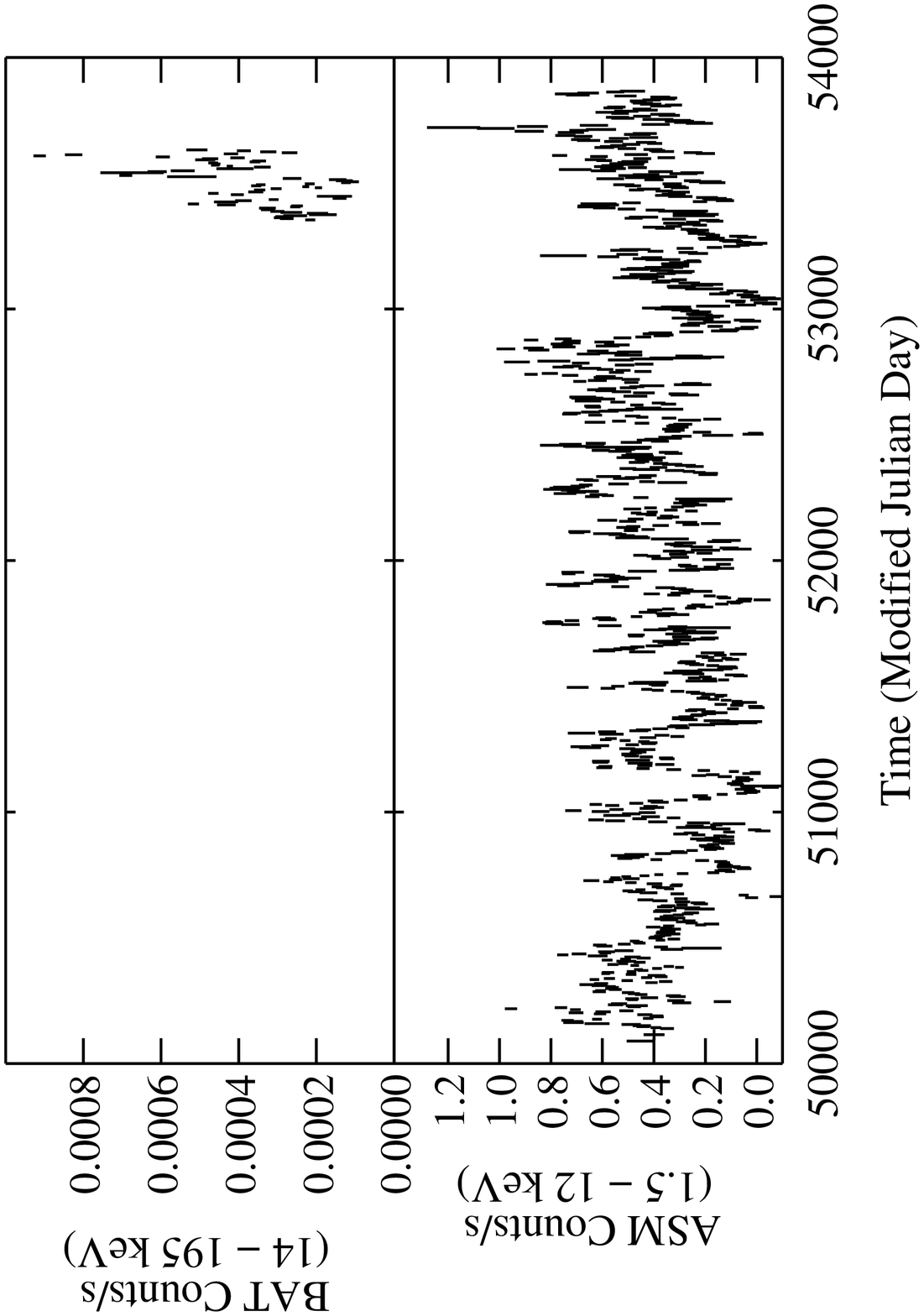]{RXTE ASM (lower panel) and Swift BAT
(upper panel) observations of \src. Both light curves
are heavily smoothed one day averages.}

\figcaption[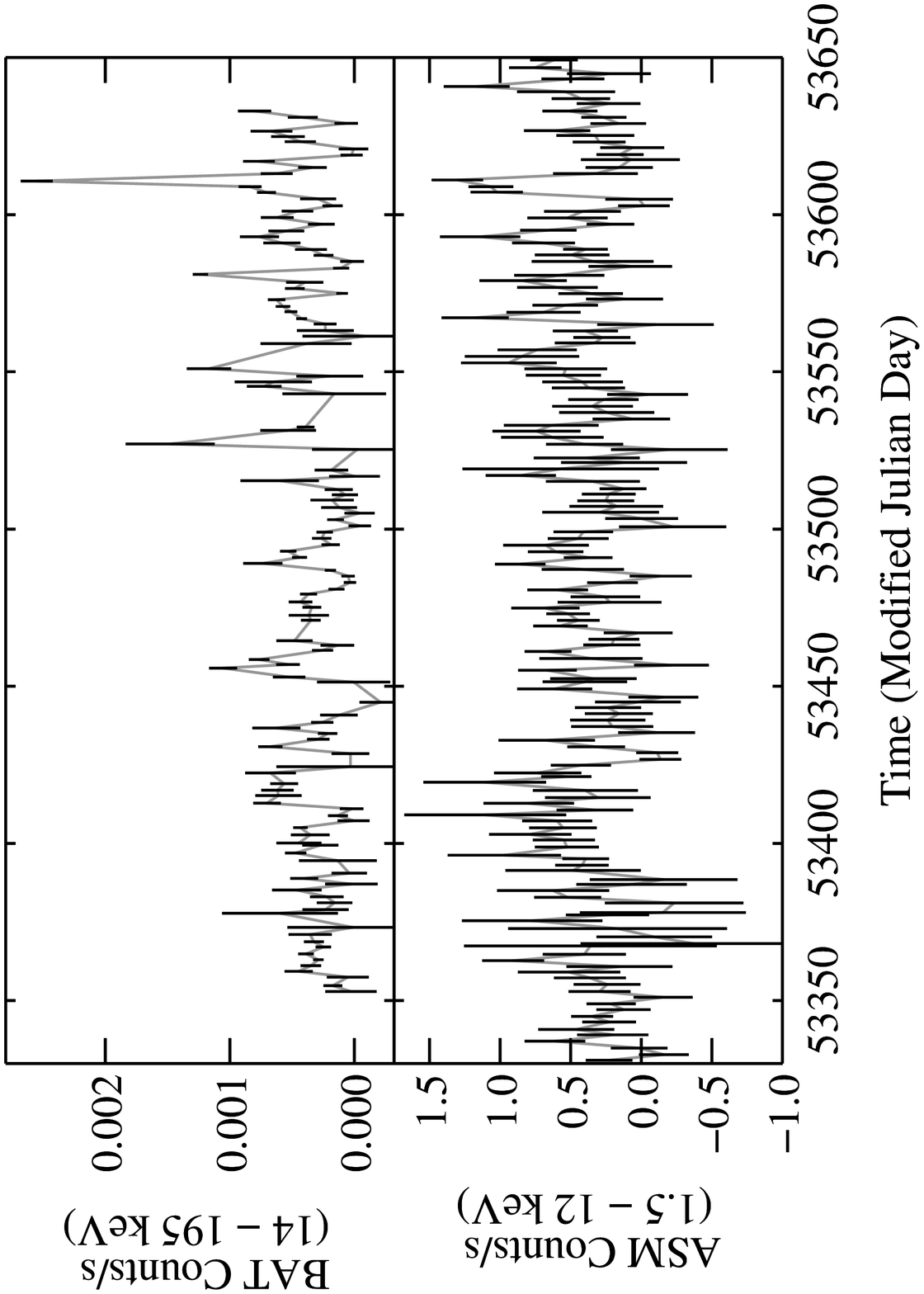]{RXTE ASM (lower panel) and Swift BAT
(upper panel) observations of \src\ from the time around
the BAT observations only. Both light curves are two
day averages with no additional smoothing. Only points which
contain at least two ``snapshots'' are plotted.}

\figcaption[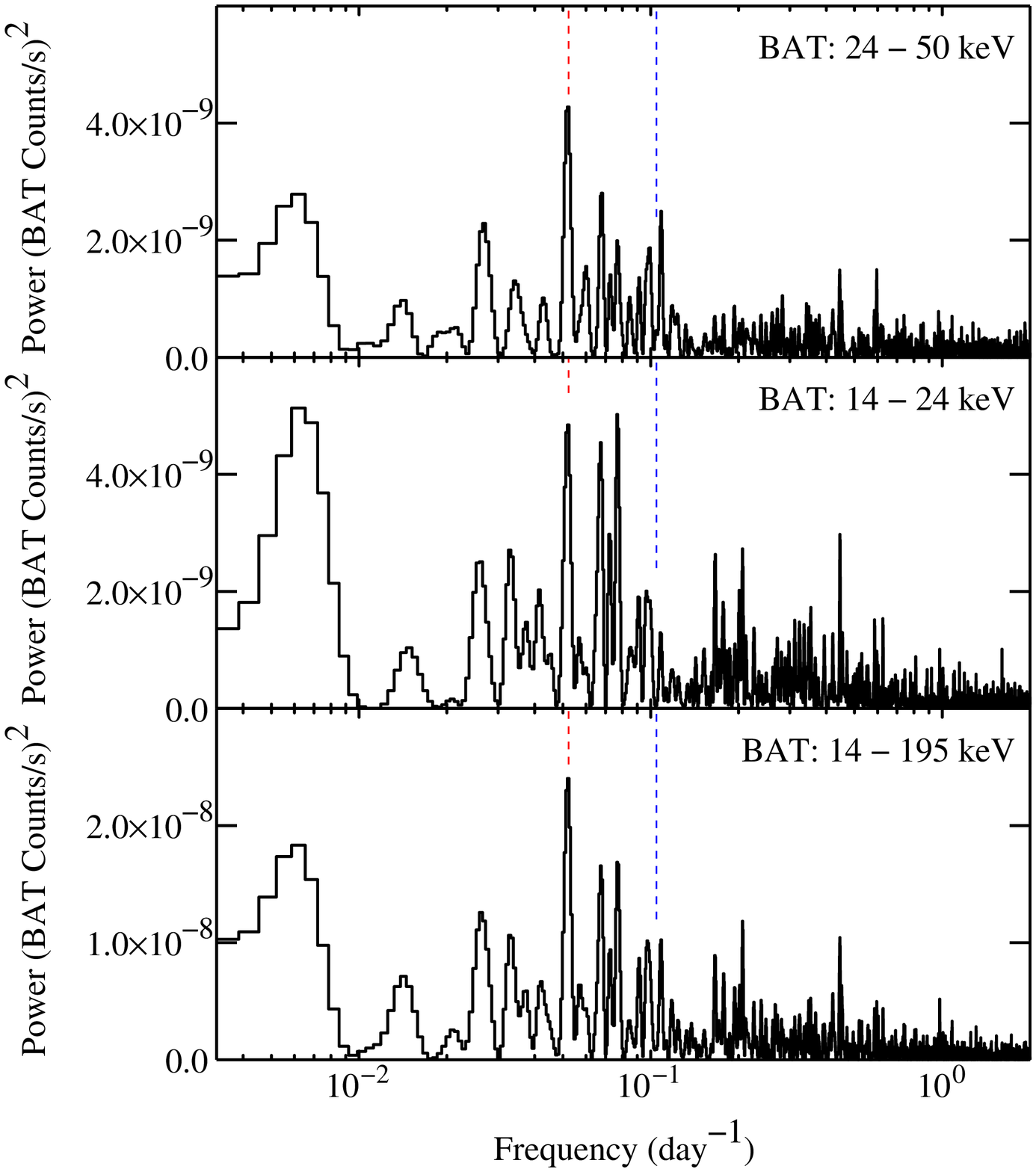]{Power spectra of BAT light curves of \src\
for the entire BAT energy range and the two lower energy range
bands. Results from the two higher energy bands are not plotted due to 
their low
count rates. The blue dashed lines indicate the 9.6 day period 
previously found
from the RXTE/ASM and the red dashed lines show twice
this period. No peak is present in any power spectrum at
the 9.6 day period.
}

\figcaption[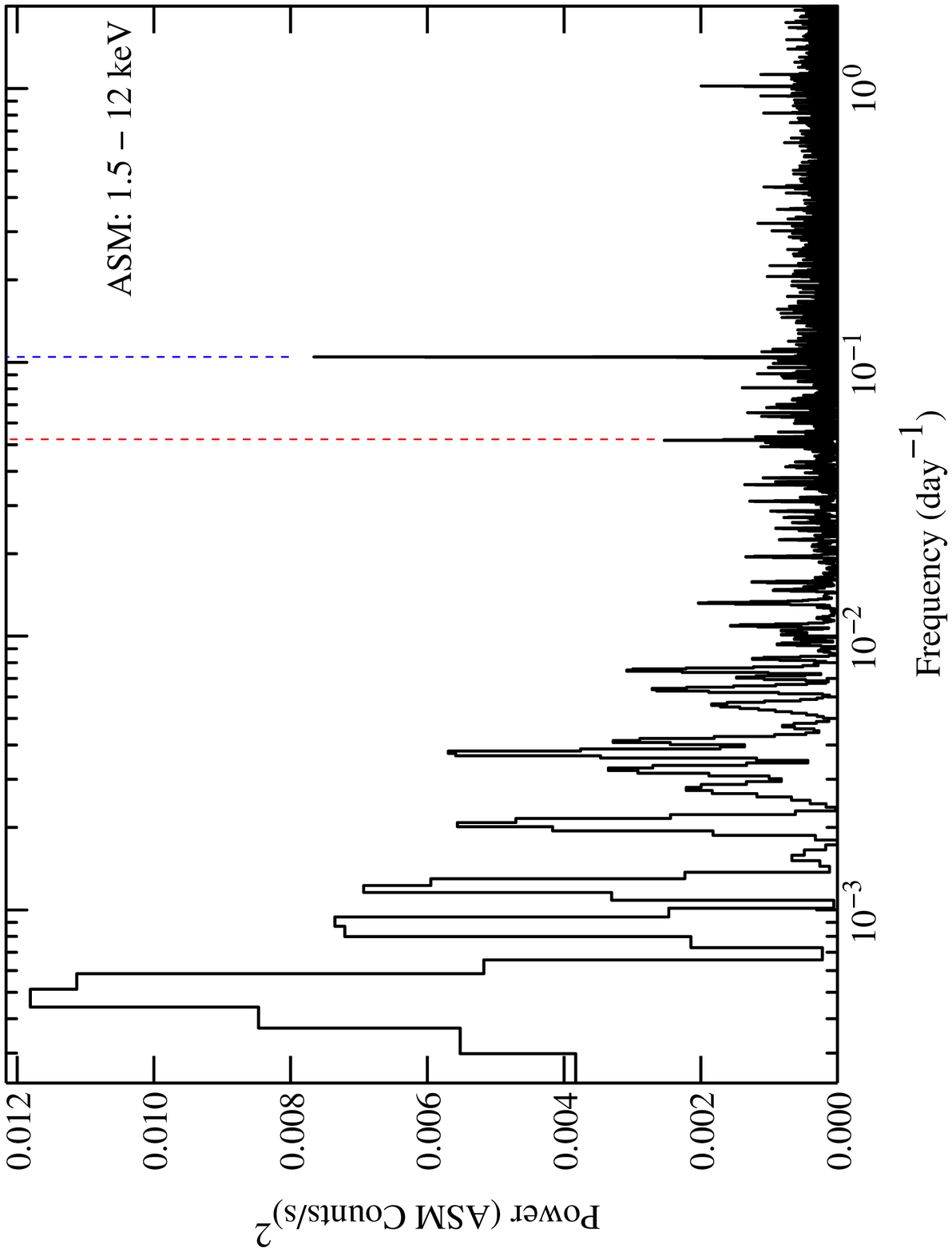]{Power spectrum of the RXTE/ASM light curve
of \src. The blue dashed line shows 
the period previously found with the ASM
and the red dashed line indicates
twice this period (``2P'').
}

\figcaption[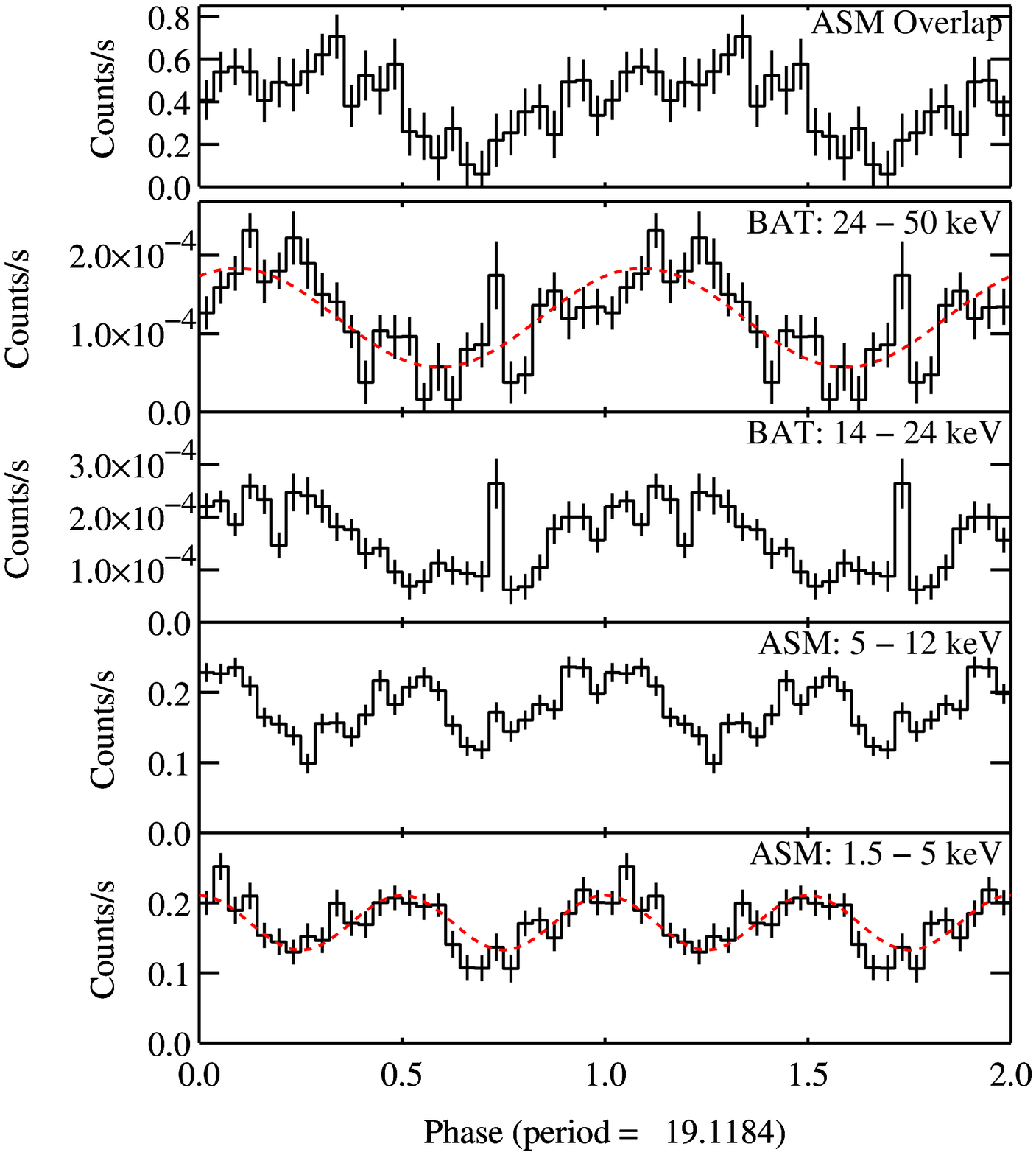]{RXTE/ASM and BAT light curves of \src\
folded on ``2P''.
The dashed lines in the top and bottom panels indicate
the results of sine wave fits to the {\em non-folded} light
curves. The upper panel shows the RXTE ASM full energy range
light curve for just the region that overlap
the BAT light curve folded on 2P.
}

\figcaption[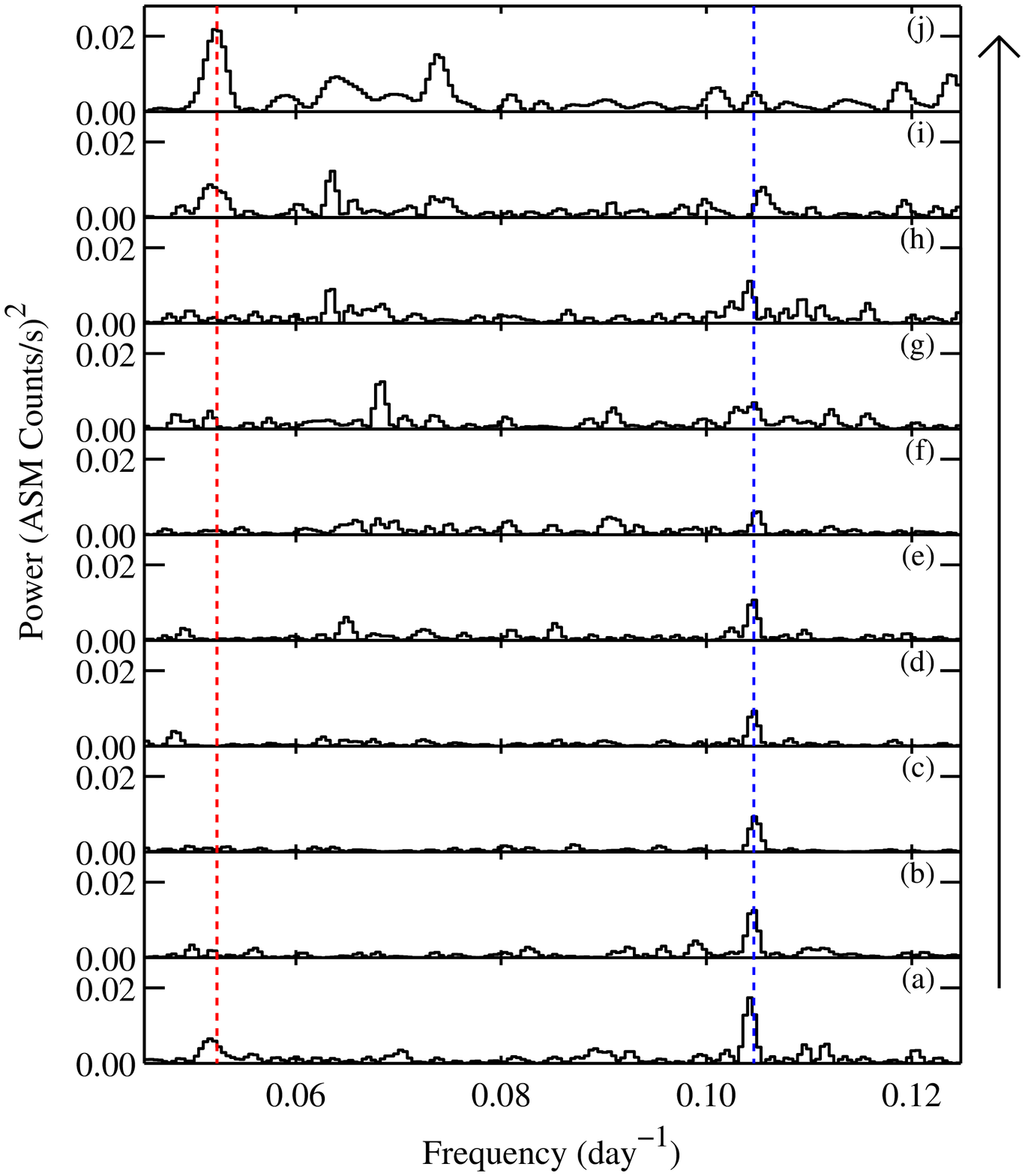]{Power spectra of the RXTE/ASM light
curve calculated from two year subsets of the data with each
subset offset by one year from the next.
The vertical dashed lines indicate ``2P'' (red) and
the previously reported 9.6 day period (blue).
The time ranges used for analysis in each section were:
 (a)
 MJD  50087 -
  50817;
 (b)
  50452 -
  51182;
 (c)
  50817 -
  51547;
 (d)
  51182 -
  51912;
 (e)
  51547 -
  52277;
 (f)
  51912 -
  52642;
 (g)
  52277 -
  53007;
 (h)
  52642 -
  53372;
 (i)
  53007 -
  53737;
 (j)
  53372
  54102.
}

\figcaption[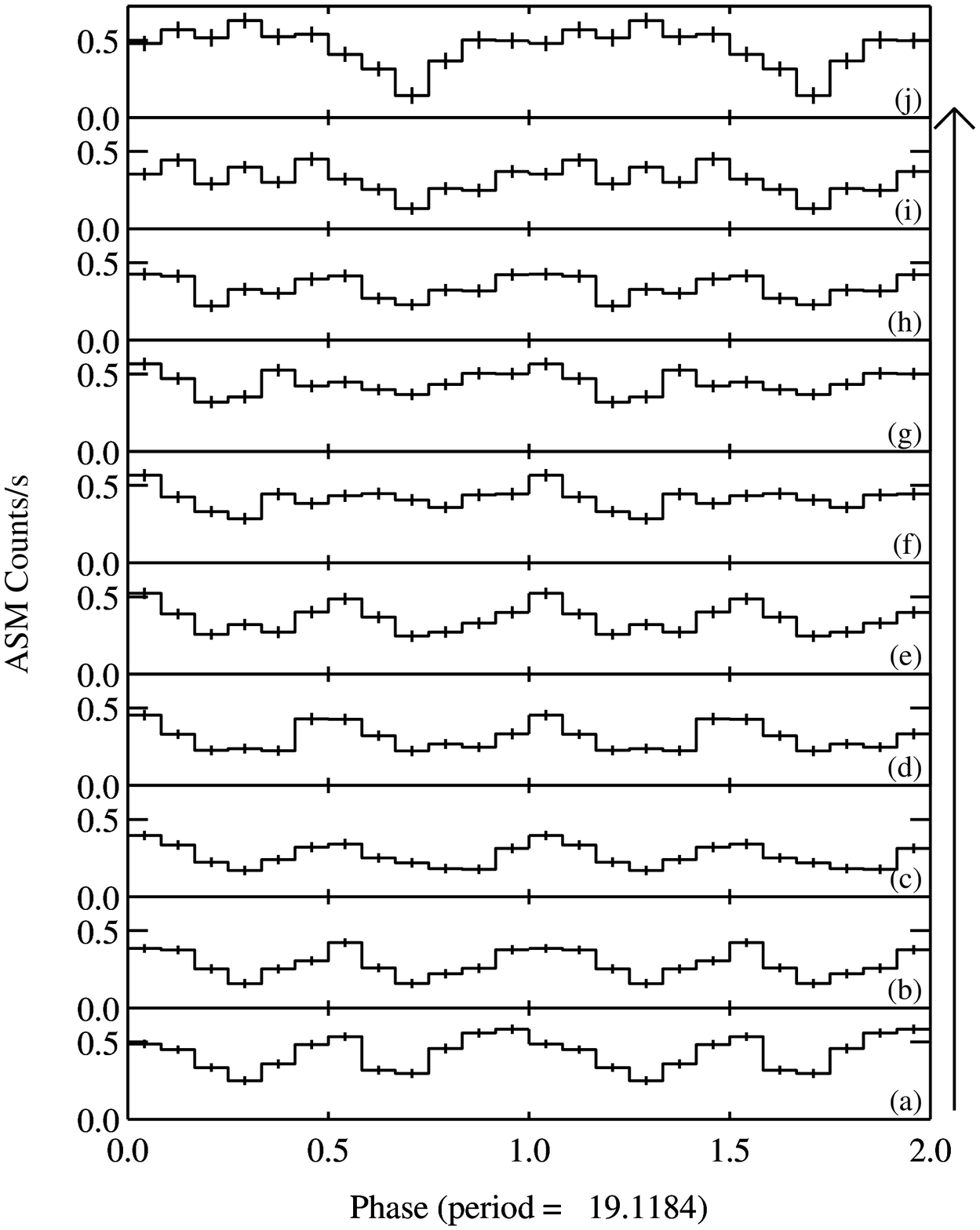]{RXTE/ASM light curves of \src\ folded
on a period of 2P. The ASM light curve was divided into
the same sections used in Fig. 7.}

\figcaption[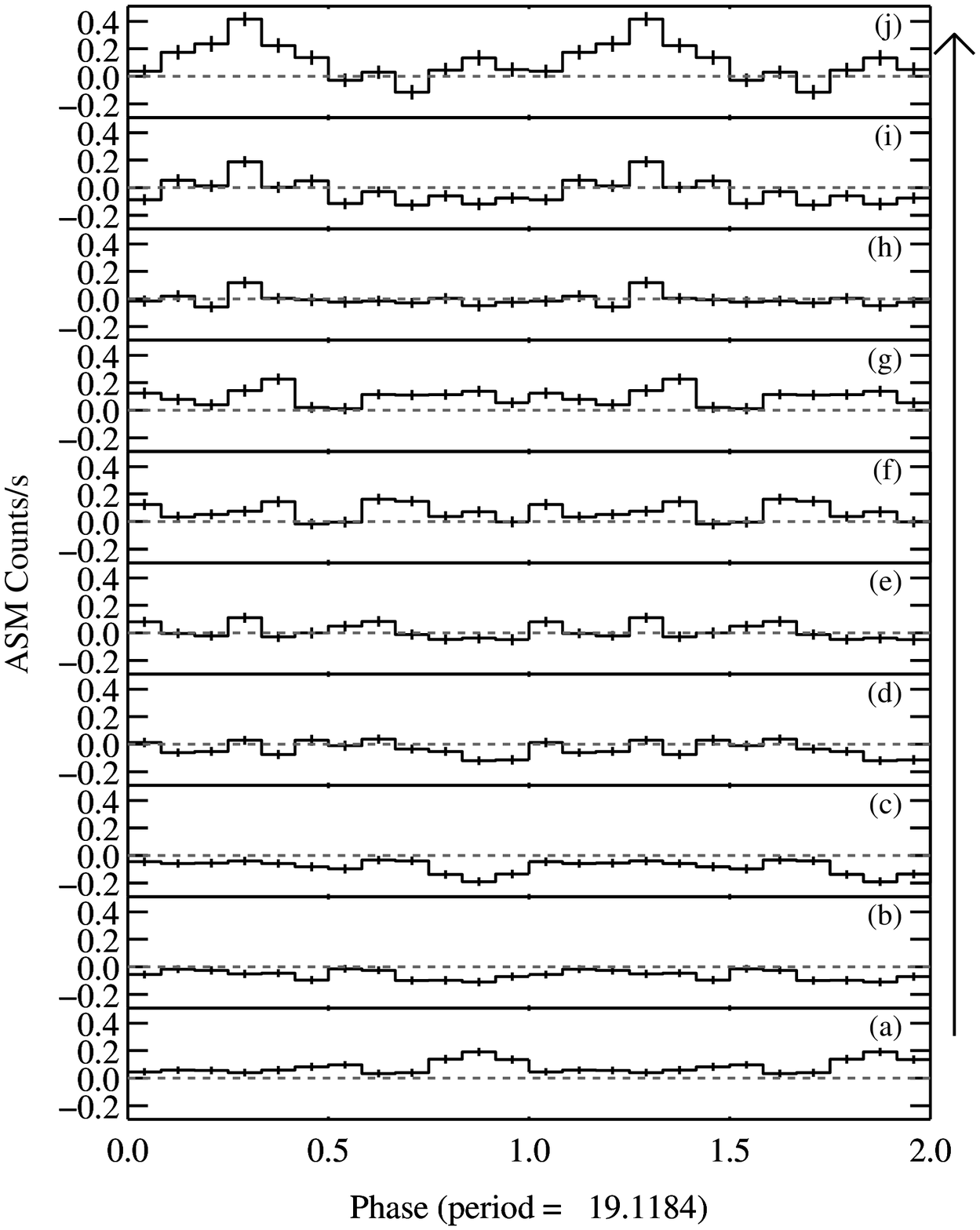]{RXTE/ASM light curves of \src\
folded on a period of 2P with the mean profile
in Fig. 8 (a) and (c) subtracted. The light curve was divided into
the same sections used in Figs. 7 and 8.}

\figcaption[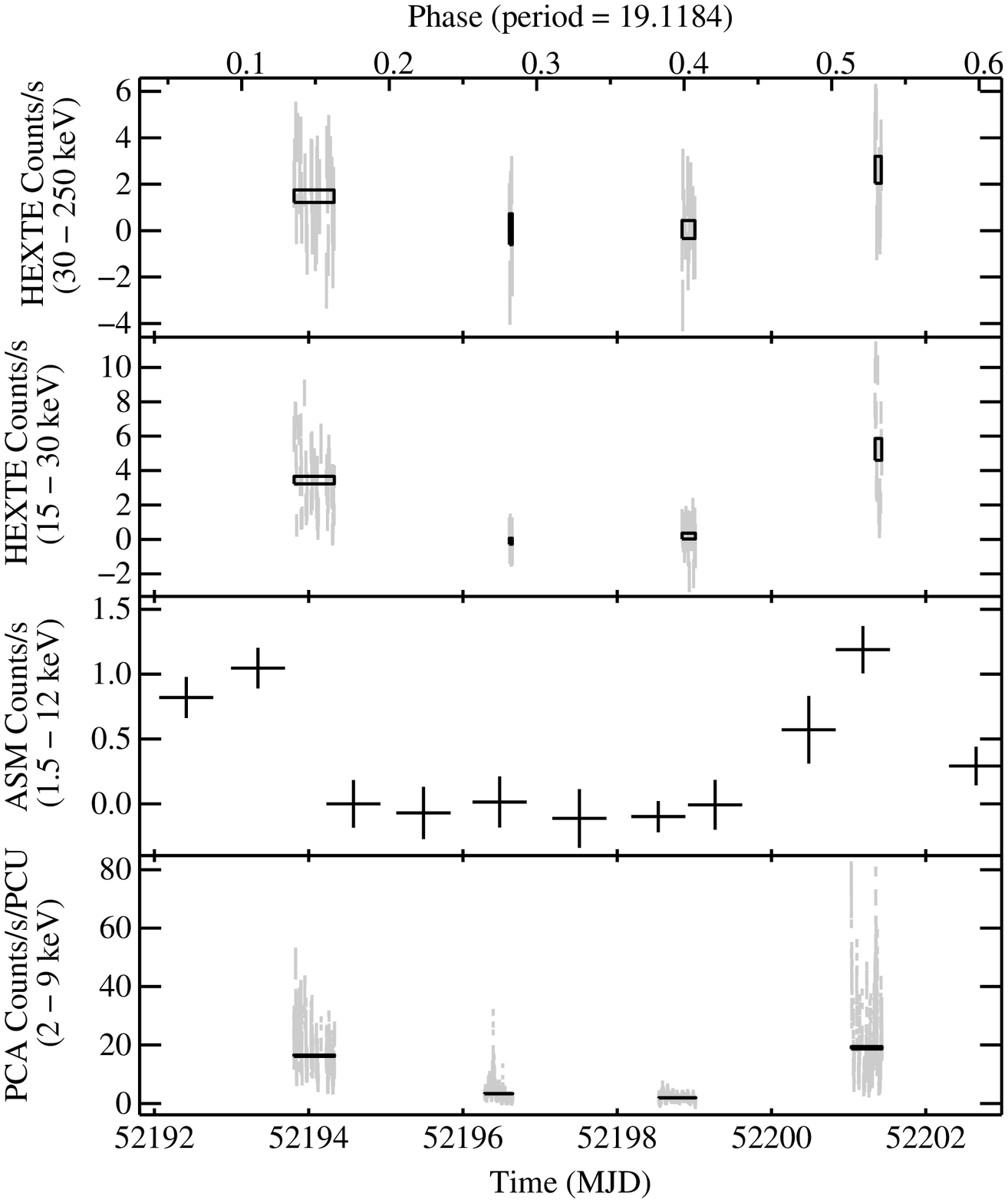]{RXTE PCA, ASM, and HEXTE light curves of \src\ obtained
in 2001. The 15 - 30 and 30 - 250 keV HEXTE light curves were 
rebinned by factors of 2 and 4 respectively compared to the original
standard products. The boxes indicate the mean 
count rate, plus and minus the standard error on this value
during each of the four PCA and HEXTE observations.
}

\begin{figure}
\plotone{f1.eps}
\end{figure}

\begin{figure}
\plotone{f2.eps}
\end{figure}

\begin{figure}
\plotone{f3.eps}
\end{figure}

\begin{figure}
\plotone{f4.eps}
\end{figure}

\begin{figure}
\plotone{f5.eps}
\end{figure}

\begin{figure}
\plotone{f6.eps}
\end{figure}

\begin{figure}
\plotone{f7.eps}
\end{figure}

\begin{figure}
\plotone{f8.eps}
\end{figure}

\begin{figure}
\plotone{f9.eps}
\end{figure}

\begin{figure}
\plotone{f10.eps}
\end{figure}

\end{document}